\documentclass{wiley-article}
\usepackage[super,square,sort&compress]{natbib}
\usepackage{changes} 
\usepackage{pdfpages}

\papertype{Research Article}
\title{Atomic-scale control of substrate-spin coupling via vertical manipulation of a 2D metal-organic framework}

\author[1,2,3]{Benjamin Lowe}
\author[2,3,4,5]{Bernard Field}
\author[2,3]{Dhaneesh Kumar}
\author[2,3]{Daniel Moreno Cerrada}
\author[1]{Oleksandr Stetsovych}
\author[2,3]{Julian Ceddia}
\author[1]{Andr\'es Pinar Sol\'e}
\author[2,3]{Amelia Dom\'{i}nguez-Celorrio}
\author[2,3]{Jack Hellerstedt}
\author[4,5]{Sin\'{e}ad M. Griffin}
\author[1]{Pavel Jel\'{i}nek}
\author[2,3]{Agustin Schiffrin}

\affil[1]{Institute of Physics of the Czech Academy of Sciences, Cukrovarnick\'{a} 10/112, 162 00 Prague, Czech Republic}
\affil[2]{School of Physics and Astronomy, Monash University, Clayton, Victoria 3800, Australia}
\affil[3]{ARC Centre of Excellence in Future Low-Energy Electronics Technologies, Monash University, Clayton, Victoria 3800, Australia}
\affil[4]{Materials Science Division, Lawrence Berkeley National Laboratory, Berkeley, California 94720, USA}
\affil[5]{Molecular Foundry, Lawrence Berkeley National Laboratory, Berkeley, California 94720, USA}

\corraddress{Benjamin Lowe, Institute of Physics of the Czech Academy of Sciences, Cukrovarnick\'{a} 10/112, 162 00 Prague, Czech Republic

Bernard Field, Molecular Foundry, Lawrence Berkeley National Laboratory, Berkeley, California 94720, USA}
\corremail{lowe@fzu.cz, BField@lbl.gov}

\begin{document}

\begin{frontmatter}
\maketitle

\begin{abstract}
Two-dimensional (2D) materials with frustrated crystal geometries can host strongly correlated electrons, potentially leading to a range of exotic many-body quantum phases such as Mott insulators, quantum spin-liquids, and Kondo lattices. The ability to control exchange-coupling within these systems is therefore highly desirable. Here, we use an atomically sharp scanning tunneling microscope probe to vertically manipulate a 2D Mott insulating honeycomb-kagome metal-organic framework (MOF) featuring Kondo-screened local magnetic moments on Ag(111). We show that by controlling the adsorption height of the MOF, we can also controllably and reversibly change the strength of Kondo coupling between the MOF's local spins and the substrate's conduction electrons. This mechanical control of Kondo coupling could be extended to other forms of interlayer exchange coupling, potentially allowing for atomic-scale design or control of spintronics technologies.

\end{abstract}
\end{frontmatter}

\section{Introduction}

The Kondo effect is a many-body phenomenon in which the spin of a magnetic impurity in a metal host is screened by the metal's itinerant electrons via scattering processes below a critical temperature, $T_\mathrm{K}$. This effect was found to be the origin of experimentally observed increases in the resistivity of metals at low temperatures.\cite{kondo_resistance_1964,kondo_effect_1968} The advent of the scanning tunneling microscope (STM) brought renewed interest in the Kondo effect.\cite{kouwenhoven_revival_2001} The ability to probe both structure and local density of states at the atomic-scale via STM has allowed for the precise study of the Kondo effect in individual magnetic impurities, such as single transition metal atoms or magnetic molecules, on metal surfaces at low temperatures.\cite{ternes_spectroscopic_2008}

The other great advantage provided by STM is the ability to manipulate single atoms and molecules using the STM tip with atomic-scale precision. Lateral manipulation of single magnetic impurities has allowed for precise studies on the influence of the environment on the Kondo effect - such as different adsorption positions of the surface and different configurations of neighboring atoms/molecules.\cite{meng_kondo_2024,iancu_manipulation_2006,neel_manipulation_2019} Lateral manipulation has also allowed for precise construction of nanostructures such as quantum corrals.\cite{crommie_confinement_1993} The influence of the modified electronic environment due to quantum confinement of substrate surface states within these nanostructures on the Kondo effect has been investigated,\cite{aapro_tuning_2024} and quantum mirage effects also observed.\cite{manoharan_quantum_2000} Vertical manipulation has also been used to study the evolution of the Kondo effect at varying impurity-surface distances.\cite{friedrich_tuneable_2024,neel_conductance_2007,vegliante_tuning_2024} This has allowed for precise tuning of Kondo coupling strength, allowing for precise measurement of parameters such as $T_\mathrm{K}$ or contributions from spin-orbit coupling.\cite{zonda_resolving_2021,hiraoka_single-molecule_2017}

Recently, the Kondo effect has also been observed in a number of two-dimensional (2D) materials on metal substrates. These materials include several transition-metal dichalcogenide (TMD) materials such as 1T-TaSe\textsubscript{2},\cite{ruan_evidence_2021} 1T-TaS\textsubscript{2},\cite{ayani_probing_2024,ayani_electron_2024} and 1T-NbSe\textsubscript{2},\cite{zhang_quantum_2024} and one metal-organic framework (MOF).\cite{kumar_manifestation_2021} The Kondo effect is not typically observed within periodic structures, as electrons (and their spins) are usually too delocalized within bands to scatter itinerant conduction electrons of the metal substrate. What unites this group of materials is that they all host a Mott insulating phase - a Coulomb repulsion-driven insulating phase in which electrons become highly localized within half-filled bands.\cite{chen_strong_2020,nakata_monolayer_2016,kim_observation_1994,lowe_local_2024} The resultant highly localized lattice of spin 1/2 electrons can couple with substrate conduction electrons to give rise to Kondo singlets. These highly correlated materials can host diverse many-body phases such as heavy-fermion Kondo lattices,\cite{ayani_probing_2024,ayani_electron_2024} or quantum spin-liquids.\cite{ruan_evidence_2021,zhang_quantum_2024,chen_evidence_2022} The ability to control the Kondo coupling strength within these systems is therefore highly desirable, and has not yet been demonstrated. 

Here, we examine variations in Kondo coupling strength of a 2D honeycomb-kagome MOF consisting of 9,10-dicyanoanthracene (DCA) molecules coordinated to copper (Cu) atoms on a Ag(111) surface. We show that local variations in MOF adsorption height lead to varying Kondo coupling strengths (measured by $T_\mathrm{K}$) as a result of different substrate-MOF hybridization. Using an STM tip to contact the MOF, we are able to controllably and reversibly alter the MOF's adsorption height via vertical manipulation, thus achieving control over the associated Kondo coupling. 

\section{Results and Discussion}

\subsection{Kondo temperature, MOF adsorption height, and MOF-substrate hybridization}

We synthesized the single-layer DCA$_3$Cu$_2$ MOF on Ag(111) following the method outlined in the literature.\cite{kumar_manifestation_2021} DCA molecules were sublimated onto the clean Ag(111) surface held at room temperature, followed by deposition of Cu atoms while the sample was cooled to $\sim$100 K. Finally, the sample was gently annealed to $\sim$200 K (see Methods for more details). This growth method maximized the yield of the well ordered DCA$_3$Cu$_2$ MOF. 

\begin{figure}[h]
    \centering    
    \includegraphics[width=0.85\linewidth]{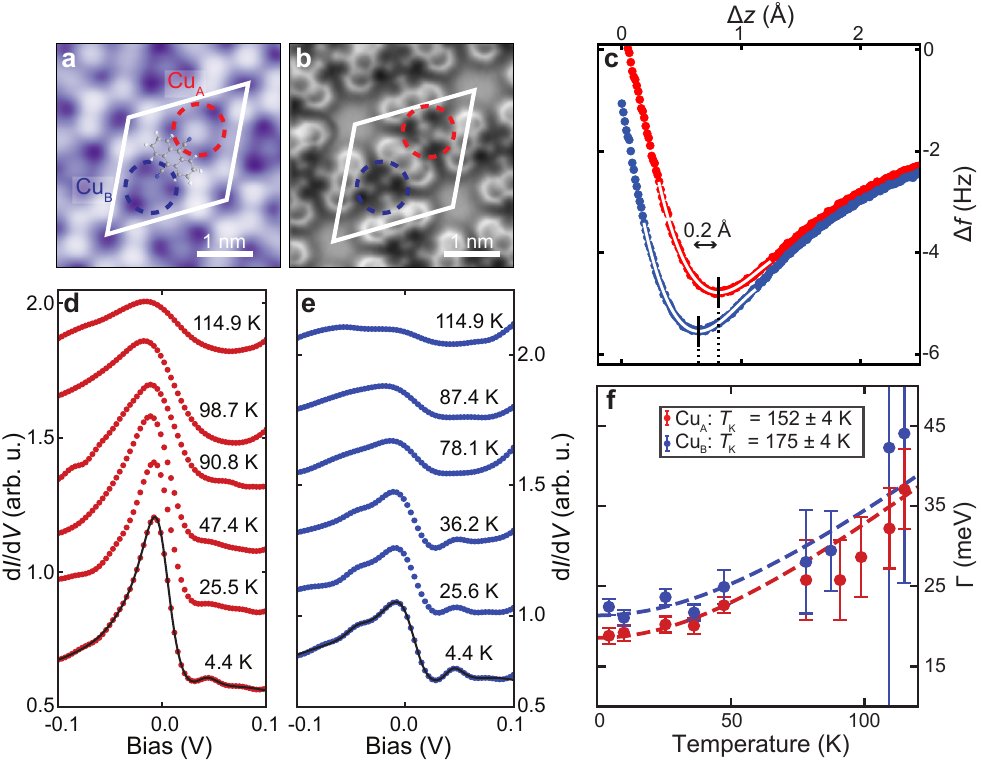}
    \caption{\textbf{Adsorption heights and Kondo temperatures of Cu sites.} \textbf{a} STM image of DCA\textsubscript{3}Cu\textsubscript{2} MOF on Ag(111) ($V_\mathrm{b} = -20$ mV, $I_\mathrm{t} = 25$ pA). DCA molecule balls-and-stick model superimposed (grey: carbon; white: hydrogen; blue: nitrogen). Dashed red (blue) circles indicate Cu\textsubscript{A} (Cu\textsubscript{B}) sites, white diamond indicates MOF unit cell. \textbf{b} NcAFM image of DCA\textsubscript{3}Cu\textsubscript{2} on Ag(111) (same region as \textbf{a}; tip functionalized with CO molecule). \textbf{c} NcAFM frequency shift, $\Delta f$, as a function of tip-sample distance $\Delta z$, acquired with a CO-tip at Cu\textsubscript{A} (red) and Cu\textsubscript{B} (blue) sites: difference in $\Delta z_{\mathrm{min}}$ -- corresponding to $\Delta f$ minimum in $\Delta f(\Delta z)$ curves -- illustrates difference of $\sim$0.2 {\AA} in adsorption height. Solid white lines: cubic fits used to determine $\Delta z_{\mathrm{min}}$. \textbf{d,e} Temperature-dependent d$I$/d$V$ spectra at Cu\textsubscript{A} and Cu\textsubscript{B} sites, respectively, illustrating the evolution of the zero-bias peak with temperature. Examples of fitted curves shown in 4.4 K data as solid black lines. A Fano function was used to capture the zero-bias peak. \textbf{f} Evolution of Cu\textsubscript{A} (red) and Cu\textsubscript{B} (blue) zero-bias peak half-width at half-maximum $\Gamma$ as a function of temperature, from which $T_\mathrm{K}$ was determined for each site. Error bars and uncertainties represent one standard deviation in fitting parameters (although minimum uncertainties of 1 meV were assigned for $T < 77$ K and 5 meV for $T > 77$ K, see SI Section 3).}
    \label{1}
\end{figure}

In this 2D MOF, DCA molecules are arranged in a kagome pattern and Cu atoms in a honeycomb pattern, with each DCA molecule coordinated (via the cyano group nitrogen atoms) to two single Cu atoms, and each Cu atom coordinated to three DCA molecules in a trigonal geometry (see Fig. \ref{1}a,b). Despite the hybrid honeycomb-kagome physical structure, the single-particle, non-interacting bandstructure of the freestanding MOF has been shown to exhibit two Dirac bands and a flat band near the Fermi energy, consistent with the prototypical kagome bandstructure.\cite{fuchs_kagome_2020,zhang_intrinsic_2016} These kagome bands have a dominant contribution from hybridization of the lowest unoccupied molecular orbitals (LUMOs) of the kagome-arranged DCA molecules, with only very slight contribution from Cu $3d$ hybridization.\cite{field_correlation-induced_2022,lowe_local_2024,kumar_manifestation_2021} Importantly, the Cu $3d$ states have been shown both theoretically and experimentally to be in a fully filled $3d^{10}$ configuration in the DCA$_3$Cu$_2$ MOF.\cite{kumar_manifestation_2021} 

Characterization of the 2D MOF on Ag(111) via STM revealed differences in the apparent heights of Cu atoms within the MOF. Figure \ref{1}a shows an STM image of a region of the MOF, with distinct bright (labeled Cu\textsubscript{A}) and dark (Cu\textsubscript{B}) Cu sites clearly visible. This difference in apparent height is also clear in a non-contact atomic force microscopy (ncAFM) image of the same region, shown in Fig. \ref{1}b, indicating that the difference between Cu sites is structural in nature (and not purely electronic). To quantify the structural difference between Cu sites, we measured ncAFM frequency shift, $\Delta f$, as a function of tip-sample distance, $\Delta z$, at both Cu\textsubscript{A} and Cu\textsubscript{B} sites. These measurements, shown in Fig. \ref{1}c, reveal a 0.2 {\AA} adsorption height difference between these sites. The distribution of these distinct Cu sites within the MOF which differ by their adsorption height does not follow a crystalline order and arises due to a lattice mismatch between the DCA$_3$Cu$_2$ MOF and the Ag(111) substrate.\cite{kumar_manifestation_2021}

Strong electron correlations within the DCA$_3$Cu$_2$ MOF give rise to localized electrons.\cite{lowe_local_2024} When the MOF is adsorbed on Ag(111), the spins of these local electrons become Kondo-screened by the substrate conduction electrons, resulting in a characteristic zero-bias peak in differential conductance (d$I$/d$V$) scanning tunneling spectroscopy (STS) measurements.\cite{kumar_manifestation_2021} The spatial distribution of the Kondo features closely follows the spatial distribution of the hybridized DCA LUMOs within the MOF, with intensity observed at the ends of the anthracene units and at the sites of the Cu-coordinated DCA cyano groups (i.e. Cu$_\mathrm{A}$ and Cu$_\mathrm{B}$ sites); see Section 2 of Supporting Information (SI). Though the bands hosting unpaired spins have little Cu character, the DCA LUMOs from which they mainly arise have significant weight at the cyano groups. The Cu centers therefore facilitate hybridization of LUMOs via coordination to cyano groups, and zero-bias d$I$/d$V$ Kondo features are observed at these sites.

Because of the adsorption height difference between Cu$_\mathrm{A}$ and Cu$_\mathrm{B}$ sites, the Kondo screening at these sites also exhibits differences. Figure \ref{1}d,e show temperature-dependent d$I$/d$V$ spectra acquired on Cu$_\mathrm{A}$ and Cu$_\mathrm{B}$ sites, respectively, with the characteristic zero-bias Kondo peak decreasing in magnitude and broadening as the temperature $T$ increases. Following the same method outlined previously,\cite{kumar_manifestation_2021} we used fits of the experimental d$I$/d$V$ spectra, with the zero-bias Kondo peak accounted for by a Fano function (see Methods and SI Section 3) with half-width at half-maximum $\Gamma$. The resulting temperature-dependent evolution of $\Gamma$ for each Cu site is shown in Fig. \ref{1}f. By fitting $\Gamma(T)$ to the equation:

\begin{equation}\label{Kondo_temp}
    \Gamma(T) = \sqrt{2(k_\mathrm{B}T_\mathrm{K})^2+(\pi k_\mathrm{B}T)^2},
\end{equation}
where $k_\mathrm{B}$ is the Boltzmann constant, we were able to extract the Kondo temperature $T_\mathrm{K}$ for each Cu site. This analysis reveals $T_\mathrm{K} = 152 \pm 4$ K for the Cu$_\mathrm{A}$ sites, and $T_\mathrm{K} = 175 \pm 4$ K for the Cu$_\mathrm{B}$ sites, implying that the Kondo screening is stronger for the Cu$_\mathrm{B}$ sites which are closer to the Ag(111) surface.

One of the most widely used models to describe the Kondo effect is the Anderson single-impurity model.\cite{anderson_localized_1961} Within this model, exchange processes between itinerant conduction electrons of the metal host and the spin of a magnetic impurity result in flips of the impurity spin through short-lived excitations that either empty or completely fill the originally half-filled impurity electronic state. These processes give rise to a many-body singlet ground state with a characteristic resonance in the density of states at the Fermi level (detectable via d$I$/d$V$ STS).\cite{ternes_spectroscopic_2008} Within this framework, $T_\mathrm{K}$, which describes the strength of the Kondo coupling between impurity spin and host conduction electrons, can be expressed as:\cite{hewson_kondo_1993,ternes_spectroscopic_2008}
\begin{equation}\label{Kondo_hybridization}
    k_\mathrm{B}T_\mathrm{K} \approx \sqrt{2\Delta\frac{U}{\pi}} \exp\left[{\frac{-\pi}{2\Delta}\left(\left|{\frac{1}{\epsilon_\mathrm{MOF}}}\right|+\left|{\frac{1}{\epsilon_\mathrm{MOF}+U}}\right| \right)^{-1}}\right],
\end{equation}
where $\epsilon_\mathrm{MOF}$ is the impurity spin binding energy (energy difference between the host Fermi level, $E_\mathrm{F}$, and magnetic impurity half-occupied state), $U$ is the Coulomb repulsion associated with the magnetic impurity half-filled state, and $\Delta = \rho_0\lvert V_\mathrm{hyb} \rvert^2$ where $\rho_0$ is the host metal density of states at $E_\mathrm{F}$ and $V_\mathrm{hyb}$ represents effective electronic hybridization between the magnetic impurity and the metal substrate. 

While our system, consisting of a periodic 2D correlated MOF coupled to Ag(111), is more complex than a collection of isolated impurities, the observed zero-bias d$I/$d$V$ Kondo resonances at different sites motivate approximating the system as isolated Kondo impurities.
Neglecting inter-site hopping within the kagome MOF is reasonable because $U$ is larger than the kagome bandwidth, which -- in the absence of a substrate -- would put the MOF in a Mott insulating state hosting localized electrons.\cite{lowe_local_2024} Moreover, it has been shown that a system of weakly coupled correlated sites can be approximated as isolated Kondo impurities (albeit with renormalized parameters).\cite{perera_spatially_2010, romero_theory_2011}
As such, despite not capturing all the many-body properties of the correlated DCA$_3$Cu$_2$ MOF, the Anderson impurity model provides a meaningful relationship between $T_\mathrm{K}$, effective $|V_\mathrm{hyb}|$, and MOF adsorption height.

The difference in $T_\mathrm{K}$ between the Cu$_\mathrm{A}$ and Cu$_\mathrm{B}$ sites in the DCA$_3$Cu$_2$ MOF can be understood within the framework of Eq. (\ref{Kondo_hybridization}). The most significant parameter to change due to the adsorption height difference of the two Cu sites is $V_\mathrm{hyb}$ (and hence $\Delta$). Arguably, $U$ could also vary between the two sites due to different screening by the Ag(111) substrate. If this was the most significant effect, however, we would expect a lower Kondo temperature at the Cu$_\mathrm{B}$ than at the Cu$_\mathrm{A}$ site, in contradiction with experimental findings. We therefore assume that any change in $U$ is negligible compared to changes in $V_\mathrm{hyb}$. 

\begin{figure}[h]
    \centering    
    \includegraphics[width=0.5\linewidth]{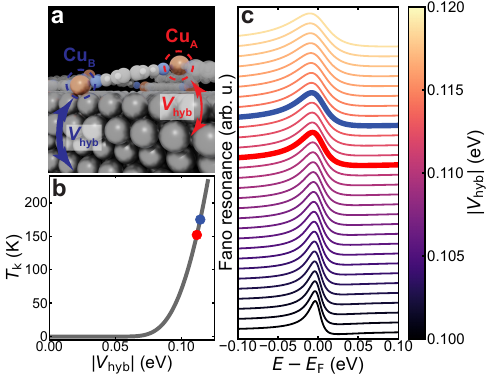}
    \caption{\textbf{Effect of MOF-substrate hybridization on zero-bias Kondo resonance and Kondo temperature.} \textbf{a} Schematic illustration of stronger MOF-substrate hybridization ($|V_\mathrm{hyb}|$) at Cu$_\mathrm{B}$ sites than at Cu$_\mathrm{A}$ sites due to MOF adsorption height difference (DCA adsorption angle within MOF exaggerated for emphasis). \textbf{b} $T_\mathrm{K}$ as a function of $|V_\mathrm{hyb}|$ according to Eq. (\ref{Kondo_hybridization}). Red (blue) marker indicates point on curve corresponding to experimental value of $T_\mathrm{K}$ for Cu$_\mathrm{A}$ site (Cu$_\mathrm{B}$ site). \textbf{c} Theoretical Fano resonance (see Methods) for different values of $|V_\mathrm{hyb}|$, using Eqs. (\ref{Kondo_temp}) and (\ref{Kondo_hybridization}) for $\Gamma$ and $T_{\mathrm{K}}$, at $T=4.4$ K (to match experiments). Highlighted red (blue) curve associated with value of $|V_\mathrm{hyb}|$ which most closely reproduces experimental spectra at Cu$_\mathrm{A}$ (Cu$_\mathrm{B}$) sites.}
    \label{2}
\end{figure}

To quantify the difference in effective impurity-host hybridization between the Cu$_\mathrm{A}$ and Cu$_\mathrm{B}$ sites, we first used Eq. (\ref{Kondo_hybridization}) to investigate the evolution of $T_\mathrm{K}$ as a function of $|V_\mathrm{hyb}|$ keeping all other parameters constant. For a detailed explanation of the parameters used for these calculations, see Methods. Figure \ref{2}b shows $T_\mathrm{K}(|V_\mathrm{hyb}|)$, with $T_\mathrm{K} \approx 0$ K for $|V_\mathrm{hyb}| \lesssim 0.07$ eV, and a steep increase in $T_\mathrm{K}$ for $|V_\mathrm{hyb}| \gtrsim 0.07$ eV. The red and blue markers in Fig. 2b indicate the points on the $T_\mathrm{K}(|V_\mathrm{hyb}|)$ evolution which match the experimentally determined $T_\mathrm{K}$ values at the Cu$_\mathrm{A}$ and Cu$_\mathrm{B}$ sites, respectively. By using Eqs. (\ref{Kondo_temp}) and (\ref{Kondo_hybridization}) for $\Gamma$ and $T_\mathrm{K}$ , we then calculated Fano functions replicating the zero-bias Kondo resonance (with $T = 4.4$ K to match experiments; see Methods), for varying $|V_\mathrm{hyb}|$ (Fig. \ref{2}c). With increasing $|V_\mathrm{hyb}|$, $T_\mathrm{K}$ and $\Gamma$ increase according to Eqs. (\ref{Kondo_hybridization}) and (\ref{Kondo_temp}), and the zero-bias peak progressively broadens. The red and blue curves in Fig. \ref{2}c highlight the values of $|V_\mathrm{hyb}|$ with $T_\mathrm{K}(|V_\mathrm{hyb}|)$ which match the experimental $T_\mathrm{K}$ values at Cu$_\mathrm{A}$ and Cu$_\mathrm{B}$ sites, respectively (and hence also match the experimental $\Gamma$ at $T$ = 4.4 K producing Fano functions which resemble the experimental zero-bias features at this temperature; Fig. \ref{1}d,e). Within this model, we found that the value of $|V_\mathrm{hyb}|$ which produce $T_\mathrm{K}(|V_\mathrm{hyb}|)$ that match the experimental $T_\mathrm{K}$ value of the Cu$_\mathrm{B}$ site is ${\sim}2 \%$ larger than that of the Cu$_\mathrm{A}$ site ($|V_\mathrm{hyb}|_\mathrm{A} = 0.112$ eV, $|V_\mathrm{hyb}|_\mathrm{B} = 0.114$ eV; although note that these effective values are likely altered by many-body effects our model does not capture - see SI Section 5). For discussion of the influence of different model parameters on this value, see SI Section 1. As shown schematically in Fig. \ref{2}a, the difference in $T_\mathrm{K}$ between the two Cu sites can be attributed to a ${\sim}2 \%$ difference in effective MOF-substrate hybridization as a result of the $\sim$0.2 {\AA} adsorption height difference.

\subsection{Atomic-scale vertical manipulation and Kondo coupling switching}

\begin{figure}[h]
    \centering    
    \includegraphics[width=\linewidth]{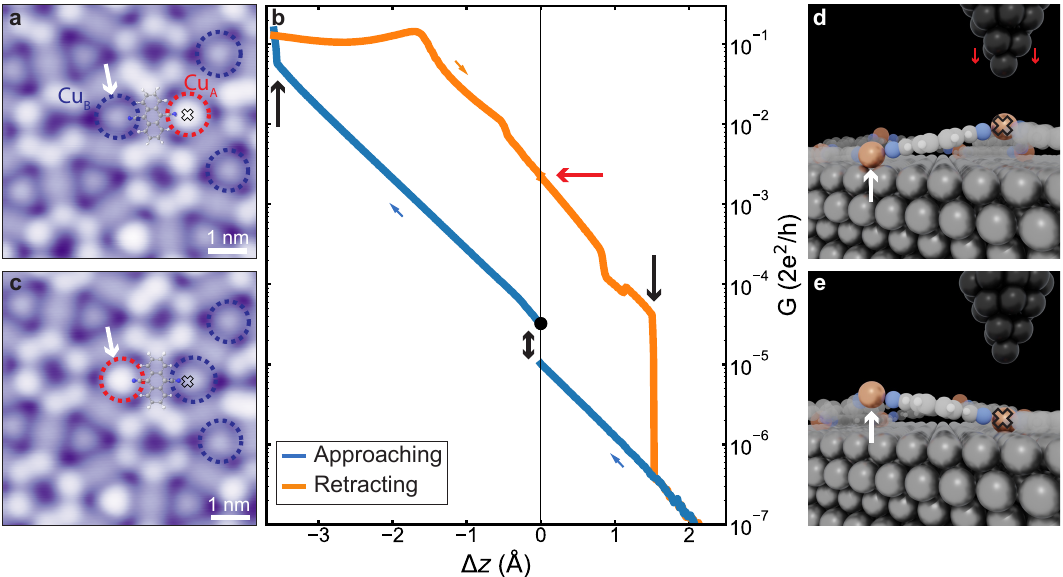}
    \caption{\textbf{Switching of Cu site adsorption height.} \textbf{a,c} STM images of DCA\textsubscript{3}Cu\textsubscript{2} on Ag(111), before and after vertical manipulation, respectively ($V_\mathrm{b} = -20$ mV, $I_\mathrm{t} = 50$ pA). DCA balls-and-sticks model superimposed. \textbf{b} Conductance, $G$, as a function of tip-sample distance difference, $\Delta z$, during vertical manipulation at site marked by cross in \textbf{a,c} ($V_\mathrm{b} = -20$ mV). Blue (orange) curve corresponds to tip approaching (retracting from) sample. Black circle indicates initial tip position at tip-sample distance defined by setpoint $V_\mathrm{b} = -20$ mV, $I_\mathrm{t} = 100$ pA. Red arrow shows increased conductance upon returning to $\Delta z = 0$. Black single-headed arrows indicate rapid changes in conductance. Black double-headed arrow indicates difference between initial and final conductance, showing switching of Cu site adsorption height. \textbf{d,e} Schematic illustration of DCA\textsubscript{3}Cu\textsubscript{2} on Ag(111) before and after manipulation with STM tip, respectively (DCA adsorption angle exaggerated for illustrative purposes). Cross and white arrow indicate corresponding positions in STM images in \textbf{a,c}.}
    \label{3}
\end{figure}

To further investigate the differences in structural and electronic properties of the two distinct Cu sites, we approached the STM tip closer to the surface, in an effort to manipulate the MOF atomic-scale structure. Figure \ref{3}a shows an STM image of a DCA$_3$Cu$_2$ MOF area, where one Cu$_\mathrm{A}$ and three Cu$_\mathrm{B}$ sites are highlighted by red and blue dashed circles, respectively. We approached the STM tip towards the surface at the Cu$_\mathrm{A}$ site, starting at a tip position $\Delta z = 0$ (black circle in Fig. \ref{3}b), while monitoring $I_\mathrm{t}$ (and hence the conductance $G=I_\mathrm{t}/V_\mathrm{B}$; see blue path in Fig. \ref{3}b). For most of this approach path, $G$ scales exponentially with changing tip-sample distance $\Delta z$. At $\Delta z \approx -3.3$ {\AA}, however, a sharp increase in conductance was observed (upward pointing black arrow in Fig. \ref{3}b). This is evidence of a snap-to-contact between the tip and 2D MOF.\cite{neel_conductance_2007,hiraoka_single-molecule_2017} We then retracted the STM tip (increasing $\Delta z$; orange curve in Fig. \ref{3}b), and observed significant deviations of $G$ from the typical exponential $\Delta z$ dependence. For a $\Delta z$ range between ${\sim}-2$ and -1.5 {\AA}, $G$ even appears to increase despite the tip retracting. This behavior indicates a chemical connection between tip and MOF, which we interpret as lifting of the MOF from the surface with the STM tip.\cite{lafferentz_conductance_2009,cahlik_light-controlled_2024} This is supported by the observation that upon returning to the original tip position ($\Delta z = 0$), $G$ was found to be two orders of magnitude larger than the original value (red arrow in Fig. \ref{3}b). From $\Delta z \approx -1$ to $1.5$ {\AA}, relatively normal exponential $\Delta z$ scaling of $G$ is observed, with some small deviations likely as a result of structural changes in the tip-MOF-Ag(111) junction. At $\Delta z \approx 1.5$ {\AA}, a sharp drop in conductance was observed, which we interpret as breaking of the tip-MOF connection, returning to the regular tunneling regime (downward pointing black arrow in Fig. \ref{3}b). After observing this, we then re-approached the tip towards the surface (blue curve), from $\Delta z \approx 2$ {\AA} to the original tip position of $\Delta z = 0$. The conductance at this original tip position was found to be smaller than its original value after this vertical manipulation (from $G_\mathrm{initial} \approx 3.2\times10^{-5}$ to $G_\mathrm{final}{\approx}1.5\times10^{-5}$ 2e$^2$/h; double-headed black arrow in Fig. \ref{3}b), indicating a change in the tunneling junction as a result of the manipulation. 

An STM image of the same sample region following the manipulation procedure is shown in Fig. \ref{3}c. The MOF appears to be relatively unchanged compared to Fig. \ref{3}a, with no sign of damage to the overall structure (and also no suggestion of changes at the tip apex). Strikingly, however, the Cu site at which manipulation was performed (black cross in Fig. \ref{3}a,c) appears to have switched from a Cu$_\mathrm{A}$ appearance to a Cu$_\mathrm{B}$ appearance. Similarly, the neighboring Cu atom to the left of the site at which manipulation was performed (white arrow in Fig. \ref{3}a,c) appears to have switched from a Cu$_\mathrm{B}$ to a Cu$_\mathrm{A}$ site. The other two Cu$_\mathrm{B}$ neighbors of the site indicated by the black cross remain unchanged by the manipulation procedure. As illustrated schematically in Fig. \ref{3}d,e, the result of the manipulation is a ``see-saw''-like flip of one pair of neighboring Cu$_\mathrm{A}$ and Cu$_\mathrm{B}$ sites connected by a DCA molecule. 

\begin{figure}[h]
    \centering    
    \includegraphics[width=\linewidth]{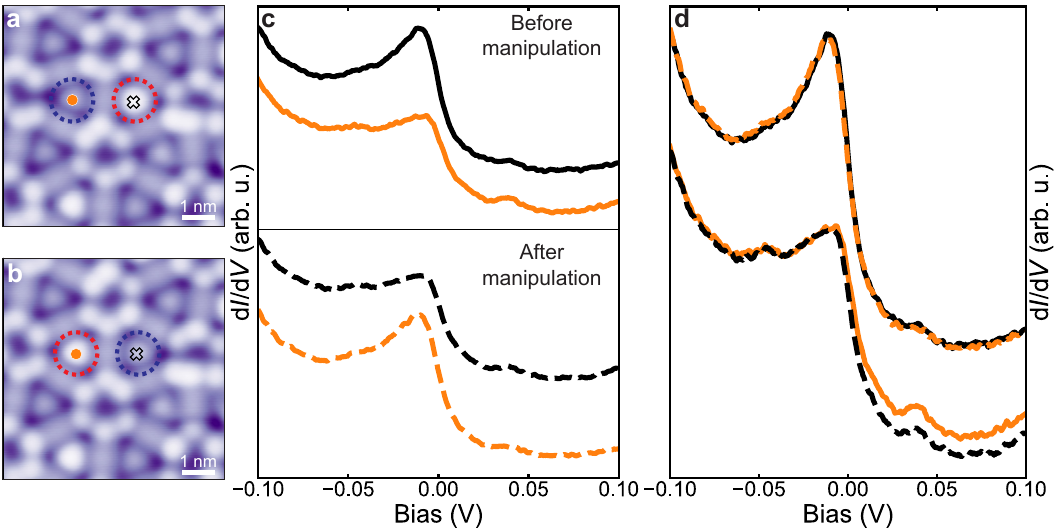}
    \caption{\textbf{Switching of Cu site Kondo coupling strengths.} \textbf{a,b} STM images of DCA\textsubscript{3}Cu\textsubscript{2}/Ag(111) before and after vertical manipulation, respectively, at initial Cu$_\mathrm{A}$ site marked by cross ($V_\mathrm{b} = -20$ mV, $I_\mathrm{t} = 50$ pA). Manipulation parameters as in Fig. \ref{3}. \textbf{c} d$I/$d$V$ spectra acquired at positions indicated in \textbf{a}, \textbf{b}, before (solid curves) and after (dashed curves) manipulation. Setpoints: $V_\mathrm{b} = -100$ mV, $I_\mathrm{t} = 100$ pA. Black curves offset for clarity. \textbf{d} Same spectra as in \textbf{c}, overlaid for comparison (the two pairs of curves are offset for clarity). Local structural and d$I$/d$V$ signatures corroborate switching from Cu$_\mathrm{A}$ to Cu$_\mathrm{B}$ site at manipulation location (black cross and spectra), and from Cu$_\mathrm{B}$ to Cu$_\mathrm{A}$ at adjacent location (orange circle and spectra).}
    \label{4}
\end{figure}

To understand the effect of the manipulation procedure on the local electronic properties of the DCA$_3$Cu$_2$ MOF, we performed d$I$/d$V$ measurements at affected Cu sites before and after manipulation. The upper panel of Fig. \ref{4}c (solid curves) shows d$I$/d$V$ spectra acquired at Cu$_\mathrm{A}$ (black) and Cu$_\mathrm{B}$ (orange) sites indicated in Fig. \ref{4}a, prior to manipulation, with zero-bias Kondo peaks associated with different  characteristic Kondo temperatures $T_\mathrm{K}$ (and hence different Kondo screening strength), as outlined in Fig. \ref{1}. The lower panel of Fig. \ref{4}c (dashed curves) shows d$I$/d$V$ spectra at the same locations, following manipulation. The d$I$/d$V$ spectrum (black solid curve) at the initial Cu$_\mathrm{A}$ manipulation site is associated with a large-magnitude, narrow zero-bias Kondo peak. This peak becomes weaker in magnitude and wider after manipulation (black dashed curve). The d$I$/d$V$ spectrum (orange solid curve) at the initial Cu$_\mathrm{B}$, adjacent to the manipulation location, is associated with a small-magnitude, wide zero-bias Kondo peak. This peak becomes larger in magnitude and narrower following the manipulation (orange dashed curve). These spectra are compared directly in Fig. \ref{4}d. The spectrum at the initial Cu$_\mathrm{A}$ manipulation site is nearly identical to the spectrum of the adjacent site following manipulation. Similarly, the spectrum at the initial Cu$_\mathrm{B}$ site (adjacent to the manipulation location; orange circle in Fig. \ref{4}a) is almost identical to the spectrum at the manipulation site (black cross) after manipulation. This is strong evidence that the observed structural changes in the DCA$_3$Cu$_2$ MOF as a result of the manipulation procedure are accompanied by switches of the effective Kondo coupling strength at the affected Cu sites.

\subsection{Reversible and extended structural and Kondo coupling switching}

The switching of Cu$_\mathrm{A}$ and neighboring Cu$_\mathrm{B}$ sites was reproducible across  several MOF unit cells, provided a sharp metal tip was used. In some cases, we observed that the manipulation affected a substantial region of the sample, and that the process was reversible. 

\begin{figure}[h]
    \centering    
    \includegraphics{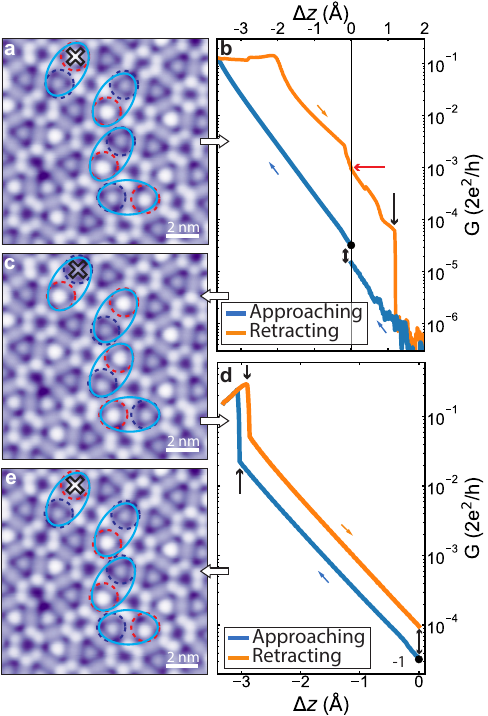}
    \caption{\textbf{Extended and reversible structural and Kondo coupling switching.} \textbf{a,c,e} STM images of DCA\textsubscript{3}Cu\textsubscript{2}/Ag(111), before and after vertical manipulation ($V_\mathrm{b} = -20$ mV, $I_\mathrm{t} = 50$ pA). Blue ovals indicate pairs of Cu sites which switch upon manipulation. Dashed red (blue) circles: Cu$_\mathrm{A}$ (Cu$_\mathrm{B}$ site) sites. \textbf{b,d} Conductance $G$ as a function of tip-sample distance change $\Delta z$, during vertical manipulation using $V_\mathrm{b} = -20$ mV, at site of black cross in \textbf{a}, \textbf{c}, \textbf{e}. Blue (orange) curves indicate tip approaching (retracting). Black circle indicates initial tip position defined by setpoint $V_\mathrm{b} = -20$ mV, $I_\mathrm{t} = 100$ pA. Red arrow shows increased conductance upon returning to $\Delta z = 0$. Black single-headed arrows indicate rapid changes in conductance. Black double-headed arrow indicates difference between initial and final conductance.}
    \label{5}
\end{figure}

A region of the DCA$_3$Cu$_2$ MOF is shown in Fig. \ref{5}a, in which four pairs of Cu$_\mathrm{A}$ and Cu$_\mathrm{B}$ sites are highlighted by light blue ovals. At the Cu$_\mathrm{A}$ site indicated by the black cross, we performed a manipulation as in Fig. \ref{3}. The evolution of $G$ as a function of $\Delta z$ during this manipulation procedure (Fig. \ref{5}b) is almost identical to that in Fig. \ref{3}b. This manipulation results in a switch of the black cross site from Cu$_\mathrm{A}$ to Cu$_\mathrm{B}$, and of its lower left neighbor from Cu$_\mathrm{B}$ to Cu$_\mathrm{A}$ (Fig. \ref{5}c). In this case, a larger region of the sample is also affected by the manipulation, with four further pairs of Cu$_\mathrm{A}$ and Cu$_\mathrm{B}$ switching. 

We then repeated the manipulation procedure at the same initial black cross site (now in a Cu$_\mathrm{B}$ state). As the tip was approached, $G$ followed the typical exponential evolution relative to $\Delta z$, until a sharp increase in $G$ was observed at $\Delta z \approx-3$ {\AA} (upward black arrow), indicating a snap-to-contact with the MOF (Fig. \ref{5}d). The tip was then retracted, with $G$ first increasing within a small $\Delta z$ range from ${\sim}-3.3$ to ${\sim}-2.8$ {\AA}, indicating lifting of the MOF. At $\Delta z \approx-2.8$ {\AA}, a sharp drop in $G$ was observed, indicating breaking of the tip-MOF connection. From there, typical exponential decay of $G$ with increasing $\Delta z$ was observed again until returning to the original tip position $\Delta z = 0$. In this case, we observed an increase in $G$ relative to the pre-manipulation value. An STM image after this second manipulation procedure is shown in Fig. \ref{5}e, where the black cross site switched back to its original Cu$_\mathrm{A}$ state. Notably, all four pairs of adjacent Cu sites affected by the original manipulation also switched back to their original configuration, with the STM image in Fig. \ref{5}a almost identical to that in Fig. \ref{5}e. The vertical manipulation of the sample at a specific location can affect an extended region of the MOF, and is a reversible process. 

It has been suggested that the DCA\textsubscript{3}Cu\textsubscript{2} MOF is a metastable phase on Ag(111) due to the lattice mismatch between MOF and substrate.\cite{kumar_manifestation_2021} The influence of local manipulation on an extended region of the MOF shown in Fig. \ref{5} is evidence of its instability, with the slightly buckled, frustrated system forced to find a new local energy minimum after being perturbed by the manipulation. This leads to a collective rearrangement and switching of several DCA molecules and pairs of Cu$_\mathrm{A}$ and Cu$_\mathrm{B}$ sites. 

In this work, we addressed the properties of different Cu sites of the 2D DCA\textsubscript{3}Cu\textsubscript{2} MOF, which is where differences in MOF adsorption height are most pronounced. It is important to note that previous work has demonstrated that the near-Fermi electronic states of the DCA\textsubscript{3}Cu\textsubscript{2} MOF -- which give rise to a Mott insulating phase at specific effective MOF electron populations, and to spins that produce Kondo features on Ag(111) -- have predominantly molecular character (with some weak Cu character due to hybridization).\cite{kumar_manifestation_2021,lowe_local_2024,field_correlation-induced_2022} The zero-bias d$I$/d$V$ Kondo resonances are observed at locations following the spatial distribution of these near-Fermi states with predominant molecular character (including at the extremities of the DCA anthracene backbone -- see SI Fig. S2).\cite{kumar_manifestation_2021} Due to the spatial extent of these molecular states which have significant weight at the cyano groups, the Cu sites -- each of which is coordinated to three DCA cyano groups -- also exhibit such zero-bias Kondo resonances in the d$I$/d$V$ spectra.\cite{kumar_manifestation_2021} 

The DCA\textsubscript{3}Cu\textsubscript{2} MOF hosts a periodic array of spins, each of which exhibits a zero-bias Kondo resonance with a high characteristic $T_\mathrm{K}$ value ($>100$ K) on Ag(111). This hybrid metal-organic system therefore meets all the criteria to host a 2D Kondo lattice.\cite{doniach_kondo_1977} In our experiments, however, we did not find any evidence for the emergence of Kondo lattice behavior.\cite{ayani_electron_2024,ayani_probing_2024} We hypothesize that the slight disorder arising from MOF-substrate lattice mismatch within the system may disrupt the fragile Kondo lattice state, reducing the critical temperature of its onset below the temperatures achievable in our experiments. 

Importantly, through the Anderson impurity model, our work links differences in experimentally measured $T_\mathrm{K}$ with changes in effective MOF-substrate electronic hybridization.
This model treats each site (Cu$_\mathrm{A}$, Cu$_\mathrm{B}$, and DCA anthracene extremities) as an independent impurity coupled to the itinerant electrons of the Ag(111) substrate. Similar approaches have been used to describe spatial variations in $T_\mathrm{K}$ for other metal-organic systems adsorbed on a metal surface. In a strongly correlated limit where hopping within the metal-organic complex (here, Cu-coordinated DCA) is small relative to $T_{\mathrm{K}}$, the system can be approximated as different, independent Kondo impurities.\cite{perera_spatially_2010,romero_theory_2011}
The application of this case to our system is an approximation. 
In our extended 2D MOF with non-zero coupling between neighboring sites and finite $U$, the values of $|V_\mathrm{hyb}|$ would likely differ from the effective values that we extract from the Anderson impurity model. This is suggested by the relatively small difference in effective $|V_\mathrm{hyb}|$ implied by our model for the $\sim$0.2 \AA{} difference in MOF-substrate height at Cu$_\mathrm{A}$ and Cu$_\mathrm{B}$ sites; prior literature suggests that $|V_\mathrm{hyb}|_{\mathrm{B}} / |V_\mathrm{hyb}|_\mathrm{A}$ should be larger (see SI Section 5).
This discrepancy is likely due to many-body effects, which can renormalize inter-site hopping and partially decouple orbitals within the MOF, and which are not captured in a single-site or non-interacting model.\cite{romero_theory_2011}
Developing a many-body model which accurately and quantitatively describes the correlated 2D MOF coupled to a metal substrate is beyond the scope of this work.
Though it cannot capture all the many-body properties of our system, the Anderson impurity model is nonetheless instructive in providing a connection between variations in $T_\mathrm{K}$, $|V_\mathrm{hyb}|$, and MOF-substrate height -- thus providing insight into atomic-scale control of Kondo coupling between a correlated MOF and a metal substrate.

While a number of STM contact measurements have been performed on individual atoms,\cite{neel_conductance_2007} molecules,\cite{zonda_resolving_2021,hiraoka_single-molecule_2017} or 1D periodic systems,\cite{lafferentz_conductance_2009,cahlik_light-controlled_2024,friedrich_tuneable_2024} to our knowledge this work is the first example of contacting a 2D MOF with an STM tip. This opens up the possibility of performing transport measurements through 2D MOFs using a precise STM tip as one electrode. 

\section{Conclusion}
We have demonstrated that variations in Kondo coupling strength of local spins in a 2D DCA\textsubscript{3}Cu\textsubscript{2} MOF on Ag(111) are due to different MOF adsorption heights and, as shown by the Anderson impurity model, varying MOF-substrate electronic hybridization. We have been able to exploit this by vertically manipulating the MOF with an STM tip to switch the adsorption height of Cu sites within the MOF, hence also switching their Kondo coupling strength. The ability to mechanically control exchange coupling without requiring a magnetic field could be translated to other 2D or layered magnetic systems, with potential applications in spintronics. The demonstrated ability to contact a 2D MOF could also be exploited for future transport measurements using an STM tip as one electrode.

\section*{Methods}
\subsection*{Sample preparation} 

The DCA\textsubscript{3}Cu\textsubscript{2} MOF was prepared on Ag(111) following a method we outlined previously.\cite{kumar_manifestation_2021} The Ag(111) crystal was cleaned by repeated cycles of sputtering with Ar$^+$ ions and annealing to $\sim$500 $^{\mathrm{o}}$C in ultrahigh vacuum (UHV). We first deposited DCA molecules (Tokyo Chemical Industry; >95\%) onto the clean Ag(111) substrate held at room temperature. Then, we cooled the Ag(111) surface to $\sim$100 K and deposited the Cu atoms. A balance of MOF domain crystallinity and size was achieved by then subsequently gently annealing the sample to $\sim$200 K. Base pressure during molecular and metal depositions was $< 5 \times 10^{-10}$ mbar.

\subsection*{Scanning probe microscopy measurements} 

All scanning probe microscopy measurements were performed at a base temperature $T=4.4$ K (unless otherwise stated) in UHV conditions (${\sim}1\times10^{-10}$ mbar) in a Createc LT-STM using a metallic Ag-terminated Pt/Ir tip. All STM images presented were acquired in constant-current mode at the specified setpoints (with bias voltage applied to the sample and tip grounded). The d$I$/d$V$ STS measurements in Fig. \ref{4} were performed using a lock-in method with a bias voltage modulation amplitude $V_\mathrm{mod} = 1$ mV (zero-to-peak) and frequency $f_\mathrm{mod} = 971$ Hz. The temperature-dependent d$I$/d$V$ measurements in Fig. \ref{1} were performed by numerically differentiating $I(V)$ datasets. In these temperature-dependent measurements, the base temperature of the system was kept at 4.4 K, with the STM then heated to a specified temperature with a built-in Zener diode. We waited for the system to stabilize and the thermal drift to settle prior to d$I$/d$V$ data acquisition. For STS measurements performed above 77 K, the STM was heated from a base temperature of 77 K. The experimental d$I$/d$V$ spectra in Fig. \ref{1}d, e were fitted following a procedure outlined in previous work,\cite{kumar_manifestation_2021} with the zero-bias Kondo resonance accounted for by a Fano function (see below), and effects of trivial thermal broadening included. Full details of the fitting method, including fit parameters extracted, are shown in Supporting Information Section 3. NcAFM measurements were performed using a qPlus sensor (resonance frequency $f_0 \approx 30$ kHz, $Q \approx 66$k, spring constant $K \approx 1.8$ kN/m, 60 pm amplitude modulation) with an Ag-terminated Pt/Ir tip that was functionalized with a carbon monoxide (CO) molecule for measurements. NcAFM images were acquired in constant-height mode. Adsorption-height differences between Cu sites were estimated by comparing the tip-sample distances, $\Delta z$, associated with the minima of the $\Delta f(\Delta z)$ curves in Fig. \ref{1}c. This  $\Delta z$ value associated with the minimum of the $\Delta f(\Delta z)$ curve was determined by fitting $\Delta f(\Delta z)$ with a cubic function within a $\Delta z$ range in the vicinity of the curve minimum (solid white lines in Fig. \ref{1}c). An open-source STM automation software package was used to assist with data collection and tip preparation.\cite{ceddia_scanbot_2024}

\subsection*{Anderson impurity model and zero-bias Kondo resonance simulation}

Using the Anderson impurity model, we plotted $T_\mathrm{K}$ as a function of $|V_\mathrm{hyb}|$ using Eq. \eqref{Kondo_hybridization} in Fig. \ref{2}b. The quoted values of $|V_\mathrm{hyb}|_\mathrm{A}$ and $|V_\mathrm{hyb}|_\mathrm{B}$ represent the values of $|V_\mathrm{hyb}|$ which resulted in $T_\mathrm{K}(|V_\mathrm{hyb}|)$ values that matched the experimental $T_\mathrm{K}$ values for Cu$_\mathrm{A}$ ($152 \pm 4$ K) and Cu$_\mathrm{B}$ sites ($175 \pm 4$ K), respectively. The percentage difference between $|V_\mathrm{hyb}|$ values was calculated using $\Delta|V_\mathrm{hyb}| = (|V_\mathrm{hyb}|_\mathrm{B} - |V_\mathrm{hyb}|_\mathrm{A})/|V_\mathrm{hyb}|_A$.

Within Eq. \eqref{Kondo_hybridization}, we chose the value of $\rho_0$ based on the known density of states of Ag(111) at $E_\mathrm{F}$ of $\sim$0.27 eV$^{-1}$ atom$^{-1}$.\cite{fuster_electronic_1990} We assumed that half of the atoms in a 7x7 Ag(111) supercell (corresponding to approximately one MOF unit cell)\cite{kumar_manifestation_2021} contributed to the Kondo coupling, to reflect the presence of two Cu sites per MOF unit cell. As discussed in detail in SI Section 1, however, different $\rho_0$ values rescale both $|V_\mathrm{hyb}|_\mathrm{A}$ and $|V_\mathrm{hyb}|_\mathrm{B}$ equally, such that there is no effect on the percentage difference between these two quantities if $\rho_0$ varies.

The values of $U$ and $\epsilon_\mathrm{MOF}$ are typically determined by the energy positions of the upper (UHB) and lower (LHB) Hubbard bands. However, we were not able to unambiguously detect these features for the MOF experimentally on Ag(111).\cite{kumar_manifestation_2021} Therefore, we chose $U = 0.65$ eV and $\epsilon_\mathrm{MOF} = -0.2$ eV based on measurements of the same MOF on hBN/Cu(111).\cite{lowe_local_2024} Note that these values possibly represent overestimates, as we expect greater screening on Ag(111) -- resulting in smaller $U$ -- and the formation of a Kondo resonance is known to renormalize the UHB and LHB -- resulting in a reduction of both $U$ and $\epsilon_\mathrm{MOF}$.\cite{ruan_evidence_2021,ayani_probing_2024,zhang_quantum_2024} The potential influence of this possible overestimate is discussed in SI Section 1.

Finally, we modeled the spectral Kondo resonances in Fig. \ref{2}c as Fano functions to provide further comparison between experimental data and the influence of $|V_\mathrm{hyb}|$ within the Anderson impurity model:\cite{ternes_spectroscopic_2008, kumar_manifestation_2021}

\begin{equation}
    f_\mathrm{Fano}(E, T) \propto \frac{(\epsilon(E, T)+q)^2}{\epsilon(E, T)^2+1}, \textrm{with }  \epsilon(E, T) = \frac{E-E_0}{\Gamma(T)}, 
\end{equation}
where $\Gamma(T)$ is related to the Kondo temperature $T_\mathrm{K}(|V_\mathrm{hyb}|)$ via Eq. (\ref{Kondo_temp}). We used $T = 4.4$ K to match experimental conditions, and $q = -3$ and $E_0 = -0.009$ eV to match parameters obtained from fits of the experimental d$I$/d$V$ spectra.\cite{kumar_manifestation_2021} The red and blue curves highlighted in Fig. \ref{2}c represent curves with the same $T_\mathrm{K}(|V_\mathrm{hyb}|)$ as the experimental $T_\mathrm{K}$ values for Cu$_\mathrm{A}$ and Cu$_\mathrm{B}$ sites, respectively (and hence also have matching $\Gamma(T=4.4 \text{ K})$ values).

All quoted uncertainties and error bars represent one standard deviation in fitted parameters.

\section*{Acknowledgments}
A.S. acknowledges funding from the Australian Research Council (ARC) Discovery Project scheme (DP240102006). B.L., B.F., D.K., D.M.C., J.C., J.H., and A.D.C. acknowledge funding support from the ARC Centre of Excellence in Future Low-Energy Electronics Technologies (CE170100039). B.L. and J.C. were supported through Australian Government Research Training Program (RTP) Scholarships. O.S., A.P.S., and P.J. acknowledge the support of the CzechNanoLab Research Infrastructure LM2023051, and project TERAFIT CZ.02.01.01/00/22\_008/0004594 supported by the Ministry of Education, Youth and Sports of the Czech Republic. B.L. acknowledges further funding support from the Czech Science Foundation (GA\v{C}R 25-16632I) and the European Union’s Horizon Europe research and innovation programme under the Marie Skłodowska-Curie grant agreement No. 101203634. Work at the Molecular Foundry was supported by the Office of Science, Office of Basic Energy Sciences, of the U.S. Department of Energy under Contract No. DE-AC02-05CH11231.

\section*{Conflict of Interest}
The authors declare no conflicts of interest.

\section*{Data Availability Statement}
The data supporting the findings of this study is available from the authors upon request.


\bibliography{Kondo_control}

\clearpage


\section*{Supplementary material}

# GRAPHICAL ABSTRACT

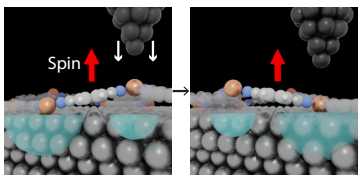

A slightly buckled 2D kagome metal-organic framework (MOF) adsorbed on Ag(111) hosts localized spins which are Kondo-screened by the underlying metal substrate. Using the tip of a scanning tunneling microscope, the MOF can be vertically manipulated to controllably change its adsorption height and with it, the Kondo coupling strength.

# Atomic-scale control of substrate-spin coupling via vertical manipulation of a 2D metal-organic framework

**Benjamin Lowe**[1,2,3]  |  **Bernard Field**[2,3,4,5]  |  **Dhaneesh Kumar**[2,3]  |  **Daniel Moreno Cerrada**[2,3]  |  **Oleksandr Stetsovych**[1]  |  **Julian Ceddia**[2,3]  |  **Andres Pinar Solé**[1]  |  **Amelia Domínguez-Celorrio**[2,3]  |  **Jack Hellerstedt**[2,3]  |  **Sinéad M. Griffin**[4,5]  |  **Pavel Jelínek**[1]  |  **Agustin Schiffrin**[2,3]

1. Institute of Physics of the Czech Academy of Sciences, Cukrovarnická 10/112, 162 00 Prague, Czech Republic
2. School of Physics and Astronomy, Monash University, Clayton, Victoria 3800, Australia
3. ARC Centre of Excellence in Future Low-Energy Electronics Technologies, Monash University, Clayton, Victoria 3800, Australia
4. Materials Science Division, Lawrence Berkeley National Laboratory, Berkeley, California 94720, USA
5. Molecular Foundry, Lawrence Berkeley National Laboratory, Berkeley, California 94720, USA

|  **Correspondence**

Benjamin Lowe, Institute of Physics of the Czech Academy of Sciences, Cukrovarnická 10/112, 162 00 Prague, Czech Republic

Bernard Field, Molecular Foundry, Lawrence Berkeley National Laboratory, Berkeley, California 94720, USA

Email: lowe@fzu.cz, BField@lbl.gov

# 1 | ANDERSON IMPURITY MODEL - PARAMETER EXPLORATION

As discussed in the main text, the values for $\epsilon_{MOF}$ and $U$ were chosen based on measurements[1] of the DCA$_3$Cu$_2$ MOF on hBN/Cu(111) measurements. Therefore, these values may have been overestimates for the same MOF on Ag(111).

Also, the value of $\rho_0$ is known to be ~0.27 eV$^{-1}$ atom$^{-1}$ for Ag(111).[2] The number of Ag atoms of the surface to consider as coupling with the MOF spins (which mostly have molecular character) was less clear, however. We considered half of the Ag atoms of the 7x7 Ag(111) supercell – corresponding to one MOF unit cell – since there are two Cu atoms within the MOF unit cell.[3]

To understand the potential influence of these parameter choices on the conclusions drawn in the manuscript, we repeated the analysis shown in Fig. 2b of the main text, for various values of $\epsilon_{MOF}$, $U$, and $\rho_0$ (Fig. S1).

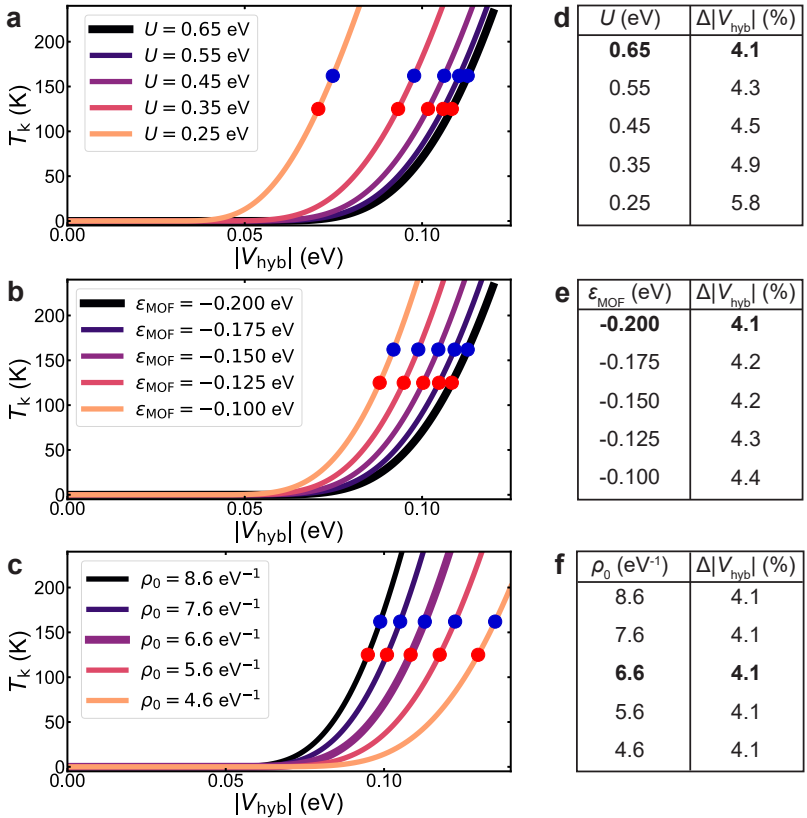

**FIGURE S1** Parameter exploration within the Anderson impurity model. **a-c** Kondo temperature $T_K$ as a function of $|V_{hyb}|$ using main text Eq. (2), with varying $U$, $\epsilon_{MOF}$ and $\rho_0$, respectively (while keeping all other parameters constant and the same as in main text Fig. 2). Red (blue) markers on each curve indicate the experimental $T_K$ values for Cu$_A$ (Cu$_B$) sites, retrieved via fitting[3] of d$I$/d$V$ spectra. Thicker curves in each panel are the same as in main text Fig. 2b. **d-f** Tables summarizing the relative difference in $\Delta|V_{hyb}| = (|V_{hyb}|_B - |V_{hyb}|_A)/|V_{hyb}|_A$. $|V_{hyb}|_A$ and $|V_{hyb}|_B$ are the values which reproduce the experimental $T_K$ according to main text Eq. (2), for Cu$_A$ and Cu$_B$ sites, respectively, for each curve in (**a-c**). Bold values indicate values used in the main text.

As can be seen in Fig. S1a-c, variations of $U$, $\epsilon_{\text{MOF}}$ and $\rho_0$ affect the evolution of $T_{\text{K}}$ relative to $|V_{\text{hyb}}|$. Smaller values of both $U$ and $|\epsilon_{\text{MOF}}|$, potentially more realistic on Ag(111) due to screening and renormalization of the MOF Hubbard bands, may lead to greater relative differences in $|V_{\text{hyb}}|$ for Cu$_A$ and Cu$_B$ sites. Variations in $\rho_0$, however, rescale both $|V_{\text{hyb}}|_A$ and $|V_{\text{hyb}}|_B$ equally by a factor of $\sqrt{\rho_0}$. Varying $\rho_0$ within the model, therefore, does not affect the percentage difference $\Delta|V_{\text{hyb}}| = (|V_{\text{hyb}}|_B - |V_{\text{hyb}}|_A)/|V_{\text{hyb}}|_A$. Overall, we conclude that the main text estimate of $\Delta|V_{\text{hyb}}| \approx 4$ % is likely a lower bound estimate, and that differences in MOF-substrate electronic hybridization between Cu$_A$ and Cu$_B$ sites caused by adsorption height differences and resulting in Kondo temperature differences may potentially be slightly larger.



\end{document}